\documentclass{ws-ijmpd}
\usepackage[super,compress]{cite}
\usepackage{xcolor}
\usepackage{graphicx}
\usepackage{epsfig}
\usepackage{psfrag}
\begin{document}

\markboth{M. M. Saez, O. Civitarese, M. E. Mosquera}{Neutrino induced reactions in core-collapse supernovae: effects on the electron fraction.}

\catchline{}{}{}{}{}

\title{Neutrino induced reactions in core-collapse supernovae: effects on the electron fraction.}

\author{M. M. Saez}

\address{Facultad de Ciencias Astron\'omicas y Geof\'{\i}sicas, University of La Plata. Paseo del Bosque S/N\\
1900, La Plata, Argentina.\\
msaez@fcaglp.unlp.edu.ar}

\author{O. Civitarese}

\address{Dept. of Physics, University of La Plata, c.c.~67\\
 1900, La Plata, Argentina\\
osvaldo.civitarese@fisica.unlp.edu.ar}
\author{M. E. Mosquera}

\address{Dept. of Physics, University of La Plata, c.c.~67\\
Facultad de Ciencias Astron\'omicas y Geof\'{\i}sicas, University of La Plata. Paseo del Bosque S/N\\
 1900, La Plata, Argentina\\
}
\maketitle

\begin{history}
\received{Day Month Year}
\revised{Day Month Year}
\end{history}

\begin{abstract}
Neutrino induced reactions are a basic ingredient in astrophysical processes like star evolution. The existence of neutrino oscillations affects the rate of nuclear electroweak decays which participates in the chain of events that determines the fate of the star. Among the processes of interest, the production of heavy elements in core-collapse supernovae is strongly dependent upon neutrino properties, like the mixing between different species of neutrinos. 

In this work we study the effects of neutrino oscillations upon the electron fraction as a function of the neutrino mixing parameters, for two schemes: the $1+1$ scheme (one active neutrino and one sterile neutrino) and the $2+1$ scheme (two active neutrinos and one sterile neutrino). We have performed this analysis considering a core-collapse supernovae and determined the physical conditions needed to activate the nuclear reaction chains involved in the r-process. We found that the interactions of the neutrinos with matter and among themselves and the initial amount of sterile neutrinos in the neutrino-sphere might change the electron fraction, therefore affecting the onset of the r-process. We have set constrains on the active-sterile neutrino mixing parameters. They are the  square-mass-difference $\Delta m^2_{14} $, the mixing angle $\sin^2 2\theta_{14}$, and the hindrance factor $\xi_s $ for the occupation of sterile neutrinos. The calculations have been performed for different values of $X_{\alpha}$, which is the fraction of $\alpha$-particles. For $X_{\alpha}=0$ the r-process is taking place if $\Delta m^2_{14} \geq 2 \, {\rm eV}^2 $, $\sin^2 2\theta_{14} < 0.8 $ and $\xi_s < 0.5$. For larger values of $X_{\alpha}$ the region of parameters is strongly reduced. The present results are compared to results available in the literature.
\end{abstract}

\ccode{ 26.30.-k,26.30.Hj,26.50.+x,26.30.Jk}

\section{Introduction}
\label{Intro}

Several experiments, like LSND \cite{aguilar01}, SK \cite{sk98}, SNO \cite{sno01}, and MiniBooNE \cite{aguilar07} among others \cite{chooz09,sage09,kamland03,gno05,k2k06,borexino08}, have provided evidences of neutrino oscillations caused by non-zero neutrino masses. In particular, the results of LSND and MiniBooNE are compatible with the inclusion of at least an extra sterile neutrino \cite{athanassopoulos95,athanassopoulos96,aguilar07b}.

The consequences of the inclusion of massive neutrinos and sterile neutrinos in different astrophysical scenarios have been examined recently \cite{boyarsky09,mohapatra04,raffelt01}. The presence of massive neutrinos affects the rates of nuclear reactions where they participate, therefore a reformulation of the weak decay rates in terms of neutrino oscillation parameters and couplings is needed to explain several astrophysical processes, such as  Big Bang Nucleosynthesis \cite{civitarese14} and nuclear reaction chains in stellar media \cite{qian93,qian95}. 
In supernovae, near the stellar-core, the neutrino flux might suffer conversions to the sterile sector causing a 
reduction in the number of electron neutrinos \cite{molinari03}. The effects of the active-active and active-sterile oscillations in supernova explosions have been studied by several authors \cite{balasi15,fetter03,balantekin04,tamborra12,wu15,janka12,pastor02,woosley94,qian03}. It allows to analyse the behaviour of matter at high densities and test properties of neutrino physics \cite{wu15,fetter03}. The process responsible for the production of nuclei heavier than iron is the rapid neutron-capture process or r-process. To be effective the reaction chain needs a neutron-rich environment, i.e. an electron fraction per baryon lower than $0.5$, a sufficiently large entropy, and short reaction times. 
The neutrino-driven matter-outflow, with a time of post-bounce of the order of 10 seconds and high neutron density, is a favoured mechanism for the production of elements heavier than iron \cite{qian03}. As said before, due to neutrino oscillations, the neutron abundance of the wind is modified when  sterile neutrinos are included \cite{qian96}.

The neutrinos determine the neutron to proton ratio, thus a possible conversion between flavours could alter this rate, and consequently the conditions needed for the r-process to take place. Calculations performed in the context of semianalytical models have shown that it is difficult to achieve the generation of the required number of free neutrons \cite{hoffman97}. The inclusion of massive neutrinos in the formalism which describes supernovae explosions affects the cross sections involved in the reaction chains that produce heavy nuclei by modifying the abundances of the elements ejected into the interstellar medium and by changing the neutrino number densities \cite{mclaughlin99,tamborra12}. In this paper, we study the impact of neutrino oscillations upon the electron fraction in the late neutrino-driven wind epoch.

This work is organized as follow. In Section \ref{estrella} we present a brief description of the supernova environment and the determination of the electron fraction. In Section \ref{densidades} we present the formalism needed to calculate the neutrino densities and show the results of the calculations in Section \ref{resultados}. In order to estimate the differences between our results and the results obtained by applying other approximations \cite{keil03,tamborra12,pllumbi15}, we have performed calculations of the electron fraction $Y_e$ as a function of the stellar radius, by using power-law and Fermi-Dirac distributions, as discussed in section \ref{resultados}. The conclusions are drawn in Section  \ref{conclusiones}.

\section{Stellar environment}
\label{estrella}
The  neutrino driven-wind is generated after the rebound of the collapsing star. For large radius, that is outside the region where the reheating mechanism takes place, the hydrostatic equilibrium is reached leading to \cite{balantekin04}
\begin{eqnarray}\label{eqp}
\frac{dP}{dr}&=&\frac{-G\rho M_{NS}}{r^2}\, ,
\end{eqnarray}
where $P$ is the hydrostatic pressure, $G$ is the gravitational constant, $M_{NS}$ is the mass of the proto-neutron star and $\rho$ is the matter density. Since the entropy at constant chemical potential is related to the pressure as $S=\frac{\partial P}{\partial T} $ one can integrate Eq. (\ref{eqp}) to obtain
\begin{eqnarray}
TS&=&\frac{G m_b M_{NS}}{r} \, .
\end{eqnarray}
In the last expression $m_b$ is the baryon mass. Near the star the radiation is the main contribution and the entropy can be written as
\begin{eqnarray}
\frac{S}{k_B}&=&\frac{2\pi^2 g_s}{45\rho_b}\left(\frac{k_BT}{\hbar c}\right)^3 \, ,
\end{eqnarray}
where $k_B$ is the Boltzmann constant, $\rho_b$ is the baryon density and  $g_s=\sum_{boson}g_b+\frac{7}{8}\sum_{fermion} g_f$ is the number of degrees of freedom \cite{balantekin04}.

Direct comparison between the last two equations yields 
\begin{eqnarray}
\rho_b &\simeq &38 \frac{2}{11} g_s \frac{M_{NS}^3}{S_{100}^4r_7^3} \, ,
\end{eqnarray}
in units of $10^3 {\rm gr\, cm}^{-3}$. $S_{100}$ represents the entropy per baryon in units of $100\, k_B$ and $r_7$ is the distance to the center of the star in units of $10^7 \, {\rm cm}$. The relationship between the entropy, radius and the temperature $T_9$ (in units of $10^9 \, {\rm K}$) is \cite{balantekin04}
\begin{eqnarray}
T_9S_{100}&\sim& \frac{2.25}{r_7} \, .
\end{eqnarray}
Different values of the entropy indicate different stages of the supernova evolution. Small values of the entropy signal a pre-heating epoch while values larger than $100\, k_B$ describe latter stages of the evolution, such as the neutrino driven-wind \cite{janka07}. We have performed the calculations with a fixed value $S_{100}=1.5$, a value which represents the late cooling phase of the neutrino-driven wind.

To characterize the neutrinos we use a Fermi-Dirac (FD)\footnote{In Section IV we have also considered a power-law distribution as suggested in Ref.  \cite{tamborra12,pllumbi15}.} distribution function \cite{balantekin04}. In the literature it is usual to assume a thermalized neutrino-flux, however the flux produced in a supernova does not necessarily has such a behaviour. The departure from thermalization is accounted for by a renormalization of the distribution function with a factor $\xi_s$. The renormalized FD distribution function will be denoted $f_{\nu}(E_{\nu})$ in the following equations. 

The flux can be obtained, after integration on solid-angles as \cite{balantekin04}
\begin{eqnarray}
\frac{d\phi_\nu}{dE_\nu}&=&\frac{c}{8\pi^2(\hbar c)^3}\frac{R^2_\nu}{r^2}E_\nu ^2 f_\nu(E_\nu) \, ,  
\end{eqnarray}
where $R_\nu$ is the radius of the neutrino-sphere. The luminosity can be computed as
\begin{eqnarray}
L_\nu&=&\frac{c R^2_\nu}{2\pi (\hbar c)^3} \int^\infty_0{E_\nu^3 f_\nu(E_\nu) dE_\nu} \, .
\end{eqnarray}
As shown in Ref. \cite{qian03,balantekin04} luminosities for different neutrino (antineutrino) species are of the order of  $10^{51}\, {\rm erg/seg}$, therefore we shall adopt this estimate in our calculations.

Weak reactions modify the amount of neutrons, protons and electrons in the star through  neutrino and anti-neutrino induced reactions
\begin{eqnarray}
\label{nu-n}
\nu_e+n &\rightarrow &p +e^- \, , \\
\label{nu-p}
\bar{\nu}_e+p &\rightarrow& n + e^+ \, .
\end{eqnarray}
The rate of these two reactions can be computed as \cite{balantekin04}
\begin{eqnarray}
\lambda_\nu&=&\int \sigma_\nu(E_\nu)\frac{d\phi_\nu}{dE_\nu }dE_\nu \, ,
\end{eqnarray}
where the cross sections, in units of ${\rm cm}^2$, are
\begin{eqnarray}
\sigma_{\nu_e}(E_{\nu_e})&=&9.6\times10^{-44}\left(\frac{E_{\nu_e}+\Delta m_{np}}{MeV}\right) \, , \\
\sigma_{\bar{\nu}_e}(E_{\bar{\nu}_e})&=&9.6\times10^{-44}\left(\frac{E_{\bar{\nu}_e}-\Delta m_{np}}{MeV}\right)\, .
\end{eqnarray}
In the last expressions $\Delta m_{np}= 1.293 \, \rm{MeV}$ is the neutron-to-proton mass-difference. Notice that for the antineutrino-cross section $\sigma_{\bar{\nu}_e}$, the antineutrino energy must be larger than the neutron-to-proton mass-difference $\left(E_{\bar{\nu}_e}>\Delta m_{np}\right)$.

The reaction rates for the inverse reactions of Eqs. (\ref{nu-n}) and (\ref{nu-p}) are written \cite{tamborra12}: 
\begin{eqnarray}
\lambda_{e^-} & \simeq & 1.578\times 10^{-2}\left(\frac{T_e}{m_e}\right)^5 e^{(-1.293+\mu_e)/T_e}
\left(1+\frac{0.646 \rm{MeV}}{T_e}+\frac{0.128 \rm{MeV^2}}{T_e^2} \right) \, ,\\
\lambda_{e^+} & \simeq & 1.578\times 10^{-2}\left(\frac{T_e}{m_e}\right)^5 e^{(-0.511-\mu_e)/T_e}\nonumber \\&&\times
\left(1+\frac{0.1.16 \rm{MeV}}{T_e}+\frac{0.601 \rm{MeV^2}}{T_e^2}
+\frac{0.178 \rm{MeV^3}}{T_e^3}+\frac{0.035 \rm{MeV^4}}{T_e^4} \right) \, .
\end{eqnarray}
In these expression $m_e$, $\mu_e$ and $T_e$ are the electron mass, the electron chemical potential and the electron temperatures (in units of ${\rm MeV}$) and the rates are given in units of ${\rm s}^{-1}$.
The electron chemical potential, at a fixed temperature, can be obtained from the equation
\begin{eqnarray}
Y_e&=&\frac{8\pi}{3N_b}T_e^2 \mu_e \left(\left(\frac{\mu_e}{T_e}\right)^2+\pi^2\right)\, ,
\end{eqnarray}
where $Y_e$ is the electron fraction.

Calling $N_j$ and $A_j$ the density and mass-number of particles of the nuclear $j$-specie, respectively, the fraction of that given specie is 
\begin{eqnarray}
Y_j&=&\frac{X_j}{A_j}=\frac{N_j}{\sum_iN_i A_i} \, .
\end{eqnarray}
If the environment is electrically neutral, the electron fraction can be computed as \cite{tamborra12}
\begin{eqnarray}
Y_e&=&\sum_j{Z_jY_j}=X_p+\frac{1}{2}X_\alpha+\sum_k\frac{Z_k}{A_k}X_k \, ,
\end{eqnarray}
where $Z_j$ is the charge of the $j$-specie. The quantities $X_p$, $X_\alpha$ and $X_k$ are the fraction mass of proton, helium and heavy nuclei, respectively.

Considering the weak reactions of Eqs. (\ref{nu-n})-(\ref{nu-p})), the proton fraction mass varies as
\begin{eqnarray}
\frac{dX_p}{dt}&=&-(\lambda_{\bar{\nu}_e}+\lambda_{e^-})X_p+(\lambda_{\nu_e}+\lambda_{e^+})X_n \, .
\end{eqnarray}

In the absence of heavy-elements the time dependence of the electron-fraction is equal to the one of protons. Also, in this case, the sum of the fraction masses of neutrons, protons and $\alpha$-particles is given by the relation $X_p+X_n+X_\alpha=1$. Taking this into account the previous equation reads
\begin{eqnarray}
\frac{dY_e}{dt}&=&\lambda_n-(\lambda_p+\lambda_n)Y_e+\frac{1}{2}(\lambda_p-\lambda_n)X_\alpha \, ,
\end{eqnarray}
where $\lambda_p=\lambda_{\bar{\nu}_e}+\lambda_{e^-}$ and $\lambda_n=\lambda_{\nu_e}+\lambda_{e^+}$.

If the plasma reaches a stage of weak equilibrium, the electron fraction does not change with time, that is $\frac{dY_e}{dt}=0$ and, therefore
\begin{eqnarray}
Y_e&=&\frac{\lambda_n}{\lambda_n+\lambda_p}+\frac{1}{2}\frac{\lambda_p-\lambda_n}{\lambda_p+\lambda_n} X_\alpha \, .
\end{eqnarray}
In the previous equation we have considered $X_{\alpha}$ as a time independent quantity. To perform the calculations we have taken different values of this quantity, starting from $X_{\alpha}=0$.


\section{Neutrino densities}
\label{densidades}

In this section we introduce the differential equation needed to compute the neutrino distribution-function as a function of the star radius. Calling $\rho$ ($\bar{\rho}$) to the neutrino (antineutrino)-distribution function in its matrix form and $\mathcal{H}$ ($\bar{\mathcal{H}}$) the neutrino (antineutrino) Hamiltonian in the flavor basis, the differential equations that give the dependence of the neutrino (antineutrino) distribution functions upon the radius are \cite{balantekin04,tamborra12}
\begin{eqnarray}
i\frac{\partial \rho}{\partial r}&=&\left[\mathcal{H},\rho\right]\nonumber \, ,\\
\label{dif}
i\frac{\partial \bar{\rho}}{\partial r}&=&\left[\bar{\mathcal{H}},\bar{\rho}\right]\, .
\end{eqnarray}

The Hamiltonian can be written as
\begin{eqnarray}
\mathcal{H}&=&\mathcal{H}^{vac}+\mathcal{H}^{m}+\mathcal{H}^{\nu-\nu} \, ,
\end{eqnarray}
where $\mathcal{H}^{vac}$ describes neutrino oscillations in vacuum, $\mathcal{H}^{m}$ represents the neutrino-matter interactions and $\mathcal{H}^{\nu-\nu}$ takes into account the neutrino-neutrino interactions. In this treatment of $\nu-\nu$ interactions we assume the single-angle approximation in which all neutrinos feel the same neutrino-neutrino refractive effect \cite{duan06,duan10}. Some works  suggest that for non-shallow matter density profiles averaging over neutrino trajectories plays a minor role in the final outcome \cite{fogli07}.


\subsection{Active-sterile neutrino mixing, $1+1$-scheme}

The active-sterile neutrino mixing in the $1+1$-scheme  is described by the Hamiltonian 
\begin{eqnarray}\label{ae11}
\mathcal{H}^{vac}_{mass}&=&pc\left(
\begin{array}{cc}
1+\frac{m_1^2c^2}{2p^2}& 0 \\
0 & 1+\frac{m_4^2c^2}{2p^2}
\end{array}
\right) \, .
\end{eqnarray}
where we have taken $E_i=\sqrt{p^2c^2+m_i^2c^4}\approx pc+\frac{m_i^2c^3}{2 p}$, $m_i$ stands for the mass of the eigenstate $i$, and $p$ is the momentum. The mixing matrix can be written as
\begin{eqnarray}\label{ae12}
U&=&\left(
\begin{array}{cc}
c_{14} & s_{14} \\
-s_{14} & c_{14}
\end{array}
\right)\, ,
\end{eqnarray}
where we have used the notation $c_{ij}=\cos \theta_{ij}$ and $s_{ij}=\sin \theta_{ij}$. The Hamiltonian in the flavour basis reads
\begin{eqnarray}
\mathcal{H}^{vac}&=&\left(pc +\frac{m_1^2c^3}{2p}\right) \left(
\begin{array}{cc}
1 & 0\\
0 & 1
\end{array}
\right)
+\frac{\Delta m^2_{14} c^3}{2p}\left(
\begin{array}{cc}
s^2_{14} & c_{14} s_{14}\\
c_{14} s_{14}& c^2_{14}
\end{array}
\right)
\, ,
\end{eqnarray}
where $\Delta m^2_{14}=m^2_4-m^2_1$.

Assuming that the sterile neutrino cannot interact with the electrons or neutrons in the star \cite{tamborra12}, the electron neutrino interactions with electrons and neutrons are described by the matter Hamiltonian. If the star is electrically neutral, the amount of electrons and protons is the same, that is $Y_e= Y_p$, and if one can neglect the presence of heavy particles the density of neutrons can be computed as $Y_n=1-Y_e$. From the definition of the electron fraction, $Y_e= \frac{N_e}{N_e+N_n}$, we can express the matter-neutrino interaction Hamiltonian as
\begin{eqnarray}
\mathcal{H}^m&=&\frac{\sqrt{2}}{2}G_f N_b \left(
\begin{array}{cc}
3 Y_e-1 & 0\\
0 & 0
\end{array}
\right) \, ,
\end{eqnarray}
where $N_b$ is the baryon density.

For the neutrino-neutrino interactions we have considered that only the $\nu_e$ and $\bar{\nu}_e$ can interact with each other \cite{tamborra12} and the terms involving sterile neutrinos vanish \cite{sigl93}, that is
\begin{eqnarray}
\mathcal{H}^{\nu-\nu}&=&\sqrt{2}G_f \left(N_{\nu_e}-N_{\bar{\nu}_e}\right)\left(
\begin{array}{cc}
2&0\\
0&0
\end{array}
\right) \, .
\end{eqnarray}
where $G_f$ is the Fermi constant and $N_{\nu_i}$ $\left(N_{\bar{\nu}_i}\right)$ is the density of the $i$-flavour neutrino (antineutrino) \cite{tamborra12}.


\subsection{Active-sterile mixing, $2+1$-scheme}

In this case we are considering three types of neutrinos in the mass-basis, a light neutrino (of mass $m_1$), a linear combination of two heavier neutrinos, the $x$-neutrino of mass $m_3$, and a sterile neutrino of mass $m_4$. The Hamiltonian $\mathcal{H}^{vac}_{mass}$ reads 
\begin{eqnarray}
\mathcal{H}^{vac}_{mass}&=&pc\left(
\begin{array}{ccc}
1+\frac{m_1^2c^2}{2p^2}& 0& 0 \\
0&1+\frac{m_3^2c^2}{2p^2}& 0 \\
0&0 & 1+\frac{m_4^2c^2}{2p^2}
\end{array}
\right) \, .
\end{eqnarray}
The mixing matrix is written as
\begin{eqnarray}
U&=& \left(
\begin{array}{ccc}
c_{14} & 0 & s_{14} \\
0 & 1 & 0\\
-s_{14} & 0 & c_{14}
\end{array}
\right) \left(
\begin{array}{ccc}
c_{13} & s_{13} & 0 \\
-s_{13} & c_{13} & 0\\
 0 & 0 & 1 \end{array}
\right)
=\left(
\begin{array}{ccc}
c_{13}c_{14} & s_{13}c_{14} & s_{14} \\
-s_{13} &c_{13} & 0 \\
-c_{13}s_{14} & -s_{13}s_{14} & c_{14}
\end{array}
\right) \, .
\end{eqnarray}
The Hamiltonian in the flavour basis reads
\begin{eqnarray}
\mathcal{H}^{vac}&=&\left(pc+\frac{m_1^2c^3}{2p}\right) \left(
\begin{array}{ccc}
1 & 0 & 0\\
0 & 1 & 0\\
0 & 0 & 1
\end{array}
\right)
+\frac{\Delta m^2_{13} c^3}{2p}\left(
\begin{array}{ccc}
c^2_{14}s^2_{13} & c_{14} c_{13}s_{13} &-c_{14} s_{14}s^2_{13}\\
c_{14} c_{13}s_{13} &c^2_{13}& -c_{13}s_{13}s_{14}\\
-c_{14} s_{14} s^2_{13}& -s_{14} c_{13}s_{13} &s^2_{14}s^2_{13}
\end{array}
\right)\nonumber \\ &&
+\frac{\Delta m^2_{14} c^3}{2p}\left(
\begin{array}{ccc}
s^2_{14} & 0& c_{14} s_{14}\\
0&0&0\\
c_{14} s_{14}& 0&c^2_{14}
\end{array}
\right)
\, .
\end{eqnarray}

The neutrino-matter interactions $ \mathcal{H}^m $ can be included as described in the previous sections
\begin{eqnarray}
\mathcal{H}^m&=&\frac{\sqrt{2}}{2}G_f N_b \left(
\begin{array}{ccc}
3 Y_e-1 & 0 & 0\\
0 & Y_e-1 & 0\\
0 & 0 & 0
\end{array}
\right) \, .
\end{eqnarray}
The neutrino-neutrino interaction-term of the  Hamiltonian  is written
\begin{eqnarray}
\mathcal{H}^{\nu-\nu}&=&\sqrt{2} G_f\left(N_{\nu_e}-N_{\bar{\nu}_e}\right)\left(
\begin{array}{ccc}
2 & 0 & 0 \\
0 & 1 & 0\\
0 & 0 & 0
\end{array}
\right) 
+\sqrt{2} G_f\left(N_{\nu_x}-N_{\bar{\nu_x}}\right)\left(
\begin{array}{ccc}
1 & 0 & 0 \\
0 & 2 & 0\\
0 & 0 & 0
\end{array}
\right) \, .
\end{eqnarray}

\section{Results}
\label{resultados}

We have solved the differential equations to calculate the neutrino (antineutrino)-distribution function Eq. (\ref{dif}) in order to obtain the electron fraction as a function of the radius of the neutrino sphere. We have performed the calculations on the late cooling phase of the neutrino-driven wind $t_{pb}\sim 10 \, \rm{sec}$. To solve the coupled differential equations we have adopted  values for the neutrino mixing parameters given in the literature \cite{meregaglia16,minakata04}. As initial condition we have taken at the neutrino sphere radius $\rm{R}_\nu=10 \, {\rm Km}$ a Fermi-Dirac neutrino distribution-function. The mean-energies were extracted from Refs. \cite{qian03,balantekin04}, namely: $<E_{\nu_e}>= 10\, {\rm MeV}$, $<E_{\overline{\nu}_e}>= 15\, {\rm MeV}$, $<E_{\nu_x}>= <E_{\overline{\nu}_x}>= 24\, {\rm MeV}$. For the sterile neutrino, we have multiplied the distribution function by a factor, $\xi_s$, which is taken as a free parameter. All the calculations have been performed assuming different values for the mass fraction of $\alpha$ particles, $X_\alpha= 0, \,\, 0.3$ and $0.5$. We have repeated some of the calculations using a power-law distribution \cite{tamborra12,pllumbi15} to compute $Y_e$. 

Active neutrinos propagate away from the SN core and convert to sterile states through MSW resonances, located at two different spacial regions. For large post bounce times inner and outer resonances are  both near the neutrino-sphere. Besides neutrino interaction with matter, neutrino-neutrino interactions also affect neutrino number densities and therefore $Y_e$. The role of neutrino self-interactions becomes more important as post bounce increases, since the matter background is lower. For later times, $\nu_e\rightarrow \nu_s$ and  $\bar{\nu_e}\rightarrow \bar{\nu_s}$ resonant conversion are expected to have the same degree of adiabaticity, resulting on a small feedback on $Y_e$ \cite{pllumbi15}.

\subsection{Active-neutrino sterile-neutrino oscillations ($1+1$-scheme)}\label{1+1}

\begin{figure}[!h]
\epsfig{figure=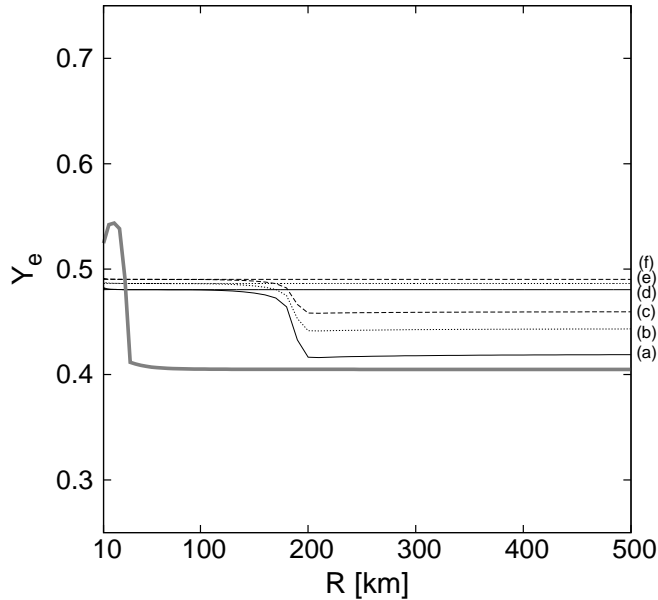, width=330pt}
\caption{$Y_e$ as a function of the radius for the $1+1$-scheme at the late cooling time. The plot was made for $\Delta m_{14}^2=2\,{\rm eV}^2$ and $\sin^2 2\theta_{14}=0.16$. The curves labelled (a), (b) and (c) stand for $X_\alpha=0$, $X_\alpha=0.3$ and $X_\alpha=0.5$ when all the interactions are considered in the Hamiltonian. The curves labelled (d), (e) and (f) stand for $X_\alpha=0$, $X_\alpha=0.3$ and $X_\alpha=0.5$ when only oscillations in vacuum are considered in the Hamiltonian. The calculations were made using a Fermi-Dirac distribution to describe the initial condition of neutrinos. The thicker line shows the results obtained by using a power-law distribution \cite{tamborra12,pllumbi15} and for $X_\alpha=0.3$.}
\label{figure1}
\end{figure}

Here we present the results obtained with the inclusion of a sterile neutrino in a core-collapse supernova and the effects of its interactions with active neutrinos upon $Y_e$. In Fig. \ref{figure1} we show the results for the fraction $Y_e$ as a function of the radius for different interactions considered in the calculations and different values of $X_{\alpha}$. The figure shows results obtained by using a FD distribution function. For comparison we have also shown a particular case calculated with a power-law distribution function and with $X_{\alpha}=0.3$ as initial condition. It can be seen that the use of a power-law distribution produces a peak at small radius and a plateau in $Y_e$ for radius larger than 20 Km.

\begin{figure}[!h]
\epsfig{figure=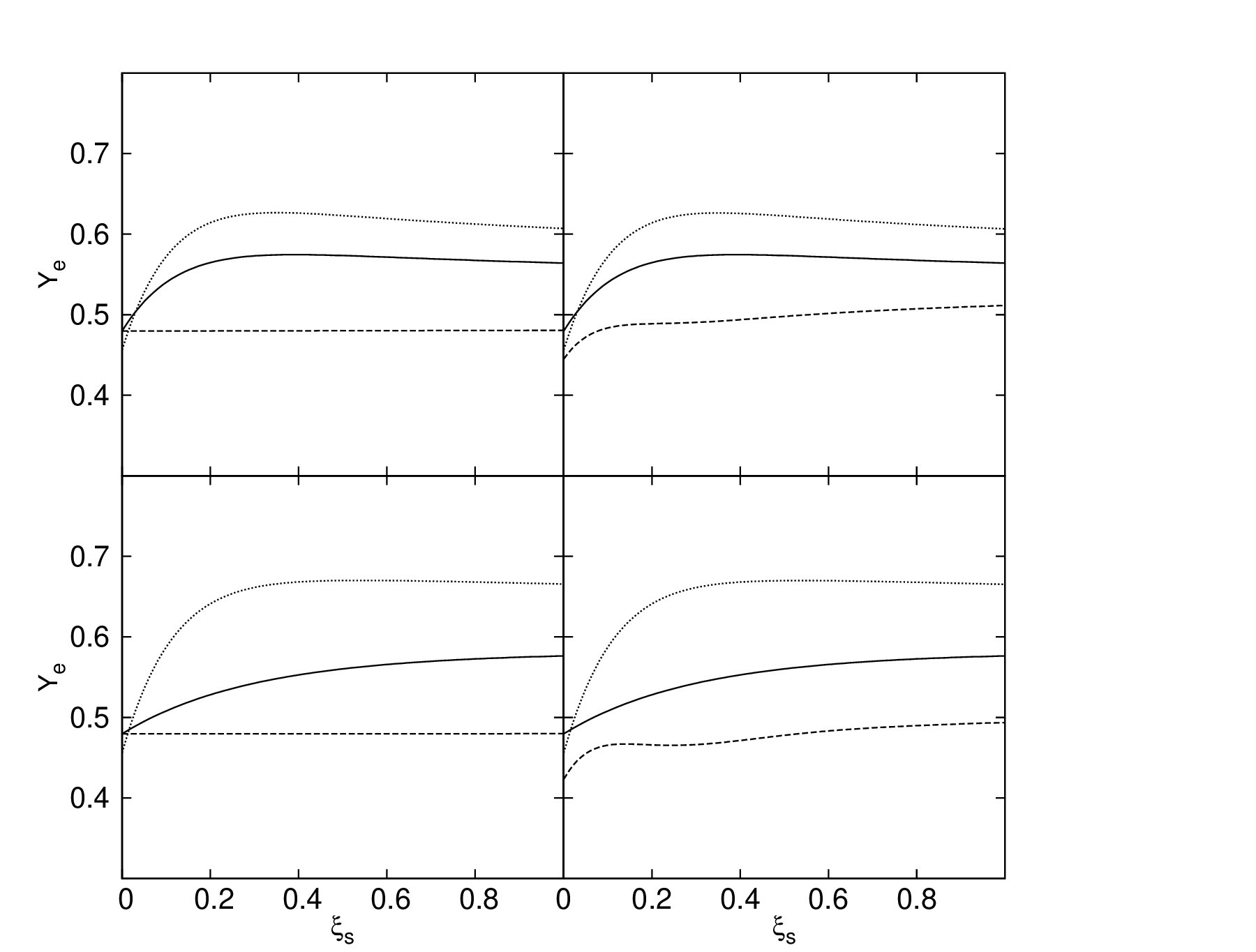,width=250 pt,angle=-90} 
\caption{$Y_e$ as a function of the initial condition $\xi_s$ for two different set of mixing parameters at $R= 75\, {\rm Km}$ (first column) and $R= 250\, {\rm Km}$ (second column) in the $1+1$-scheme. Solid line: only active-sterile neutrino oscillations; dotted line: oscillations and neutrino-matter interactions; dashed-line: oscillations, neutrino-matter and neutrino-neutrino interactions. Top row: $\sin^2 2\theta_{14}=0.5$; bottom row: $\sin^2 2\theta_{14}=0.1$. In all figures we assume normal hierarchy  for the sterile squared-mass-difference $\Delta m^2_{14}=2 \, {\rm eV}^2$, implying that the neutrino mass eigenstate $\nu_4$ is heavier than the mass eigenstates of the active neutrinos.}
\label{figure2}
\end{figure}

The calculations were performed by varying the parameter $\xi_s$ in the interval $0 \leq \xi_s \leq 1$, the square mass difference $\Delta m^2_{14}$ and the mixing angle between active and sterile-neutrinos, denoted as $\theta_{14}$. In Fig. \ref{figure2} we present the results of the electron fraction  as a function of the parameter $\xi_s$, for two different mixing angles, $\sin^2 2\theta_{14}=0.5$ and $\sin^2 2\theta_{14}=0.1$, square-mass-difference $\Delta m^2_{14}=2 \, {\rm eV}^2$, and for two radii.
When the value of $\xi_s$ is close to unity the electron fraction reaches a value larger than $0.5$. This indicates that, in order to enhance the r-process, in the neutrino-sphere should not be sterile-neutrinos. The results are in line with the findings of Ref. \cite{nunokawa97} where the authors stated that, for the considered mass range of $\Delta m_{14}^2$, conversions to sterile neutrinos in the inner core can be neglected.

\begin{figure}[!h]
\epsfig{figure=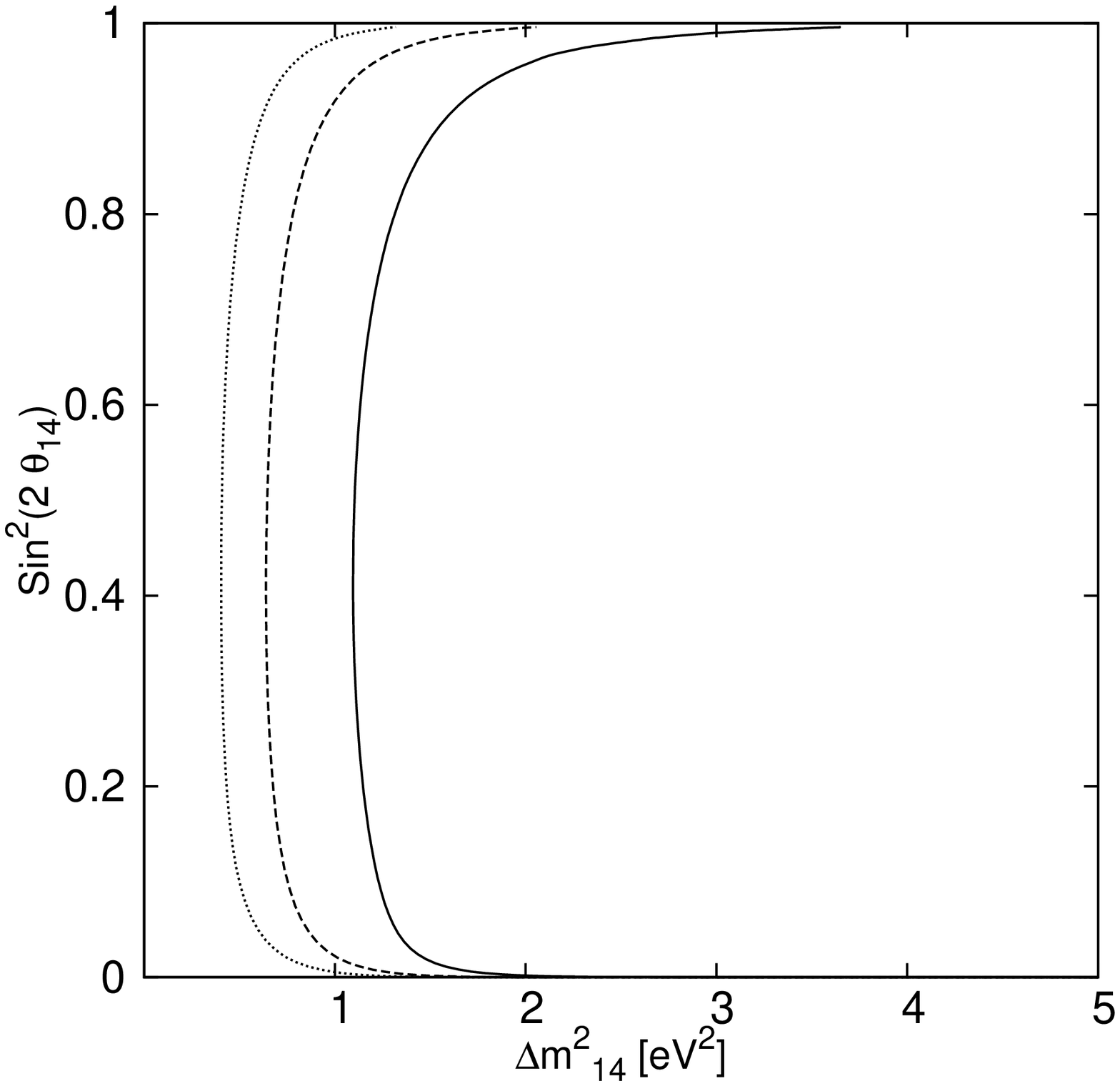,width=250 pt,angle=-90} 
\caption{Active-sterile mixing parameters for different radii in the $1+1$-scheme and for $X_\alpha=0$. The lines stand for $Y_e=0.48$ as a limiting value, calculated using all the interactions in the neutrino sector and for different radii: solid line: $\rm{R}= 150\, {\rm Km}$; dashed line: $\rm{R}= 200\, {\rm Km}$; dotted line: $\rm{R}= 250\, {\rm Km}$. The values of the parameters for which the r-process is favoured are determined by the region at the right-side of each curve, since for these regions $Y_e < 0.48$.} 
\label{figure3}
\end{figure}

To determine the allowed values of the mixing parameters we have set an upper limit for the value of the electronic fraction, $Y_e=0.48$, and varied the parameters with this constraint. In Fig. \ref{figure3} we show the constrains on the mixing angle and the neutrino's square-mass-difference for $\xi_s=0$, and three different radii, keeping $X_\alpha=0$. The curves represent $Y_e=0.48$, and the allowed regions correspond to $\Delta m_{14}^2 > 1.5 \, {\rm eV}^2$ for $\rm{R}= 150 \, {\rm Km}$, $\Delta m_{14}^2 > 1 \, {\rm eV}^2$ for $\rm{R}= 200 \, {\rm Km}$ and $\Delta m_{14}^2 > 0.5 \, {\rm eV}^2$ for $\rm{R}= 250 \, {\rm Km}$. The comparison between our results and those of Ref. \cite{nunokawa97}, for the case of the $1+1$ scheme and later epoch, shows that the set of parameters where the r-process is favoured are common to both results. The region is determined by the limits $0.5\,{\rm eV}^2<\Delta m_{14}^2 <10^2 \,{\rm eV}^2$. For larger values of $X_\alpha$ we have found that $Y_e > 0.5$, therefore suppressing the formation of heavy nuclei via r-process.

\begin{figure}[!h]
\epsfig{figure=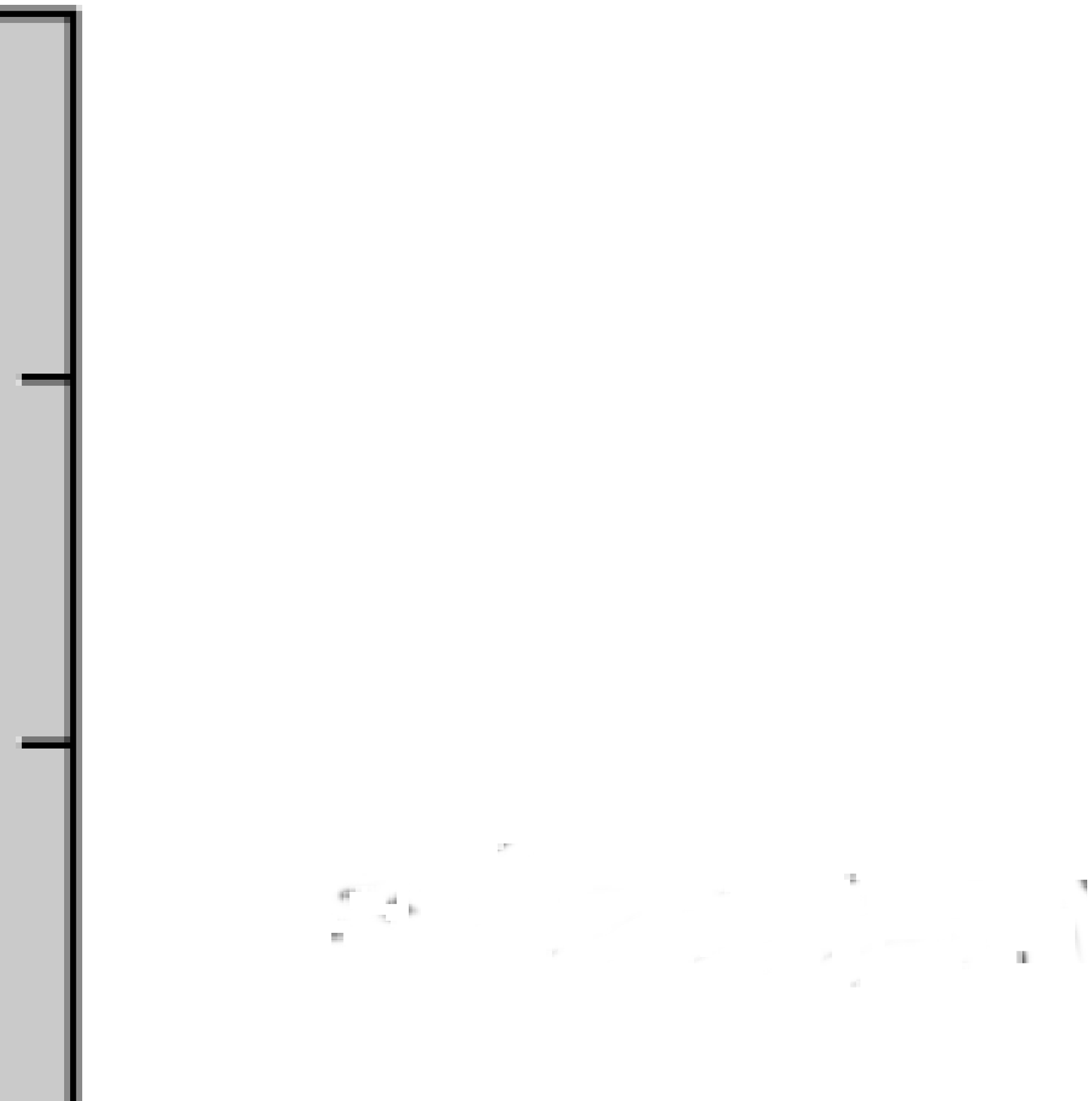,width=400 pt,angle=-180} 
\caption{Values of $\Delta m_{14}^2$ and $\xi_s$ for two different values of the mixing angle $\theta_{14}$ and $R= 150\, {\rm Km}$ in the $1+1$-scheme. First row corresponds to  $\sin^2 2\theta_{14}=0.1$ and second row for $\sin^2 2\theta_{14}=0.5$. First, second and third columns stand for $X_\alpha=0$, $0.3$ y $0.5$ respectively. White regions represent the combinations of parameters which favour the r-process ($Y_e< 0.48)$, gray regions correspond to the combinations of parameters that generate $Y_e > 0.48$.} 
\label{figure4}
\end{figure}

In Fig. \ref{figure4} we show the allowed values of $\Delta m_{14}^2$ and $\xi_s$ (white regions) for different values of $X_\alpha$, using all the interactions in the calculation of the electronic fraction, for a neutron-rich environment. The values that yield $Y_e<0.48$ are consistent with  small values of $\xi_s$. If the mixing angle increases its value the factor $\xi_s$ must decrease in order to ensure the effectiveness of the r-process. 

\begin{figure}[!h]
\epsfig{figure=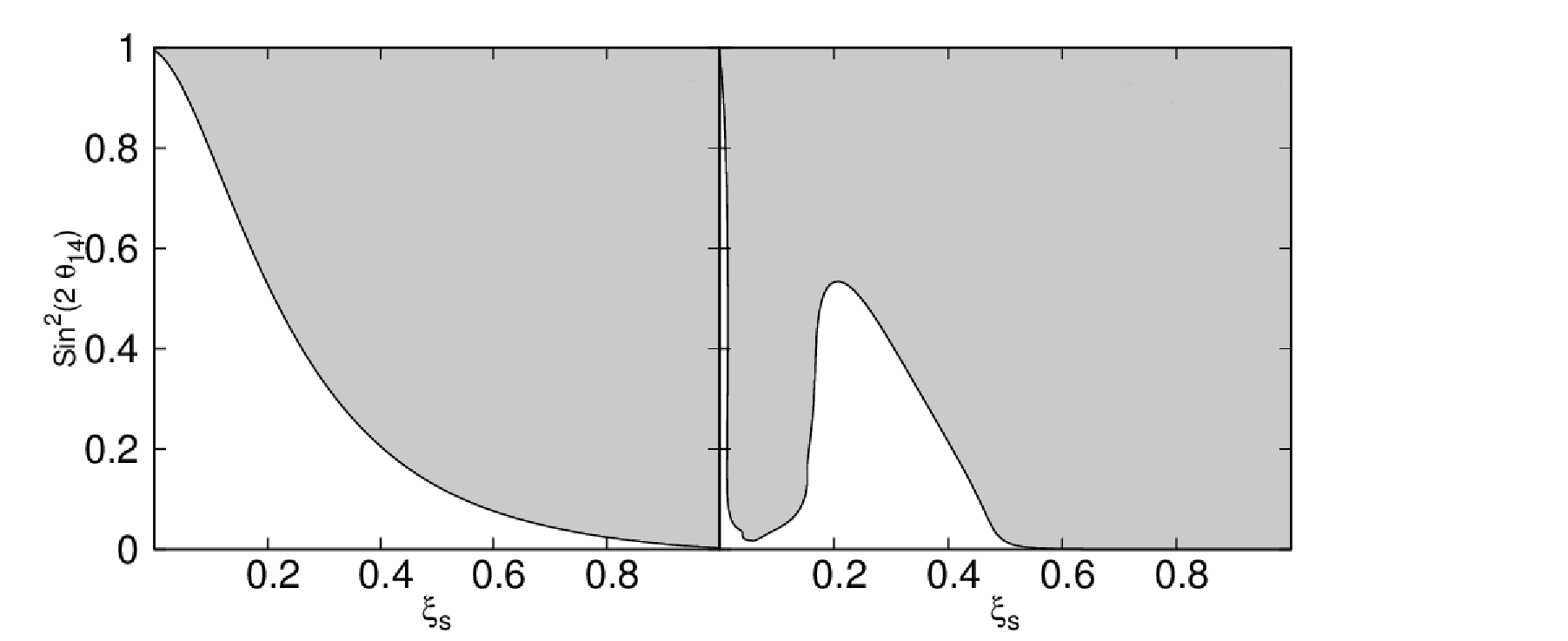,width=400 pt,angle=-180} 
\caption{Values of $\sin^2 2\theta_{14}$ and $\xi_s$ for two different values of the  square-mass-difference  $\Delta m_{14}^2$
 and  $\rm{R}= 150\, {\rm Km}$ ($1+1$-scheme). Left figure: $\Delta m_{14}^2 =2 \, {\rm eV}^2$; right figure: $\Delta m_{14}^2 = 10 \, {\rm eV}^2$. White regions represent the combinations of parameters that favours the occurrence of the r-process, while gray regions correspond to the combinations of parameters that generate $Y_e > 0.5$. The plot was made for $X_\alpha=0$. Larger values of $X_\alpha$ yields $Y_e > 0.5$}
\label{figure5}
\end{figure}

The allowed values for the mixing angle and the normalization constant of the sterile neutrino distribution function are presented in Fig. \ref{figure5} (white regions), for two different values of the square-mass-difference. Once again, the curves correspond to the value $Y_e=0.48$ when all the neutrino interactions are included in the calculation. The region which yields $Y_e<0.48$ favours small values of $\xi_s$ while the mixing-angle between active- and sterile-neutrinos is limited by the condition $\sin^2 2\theta_{14} <0.6$ when $\Delta m_{14}^2>2\, {\rm eV}^2$.

\subsection{Active-neutrino sterile-neutrino oscillations ($2+1$-scheme)}\label{2+1}

\begin{figure}[!h]
\epsfig{figure=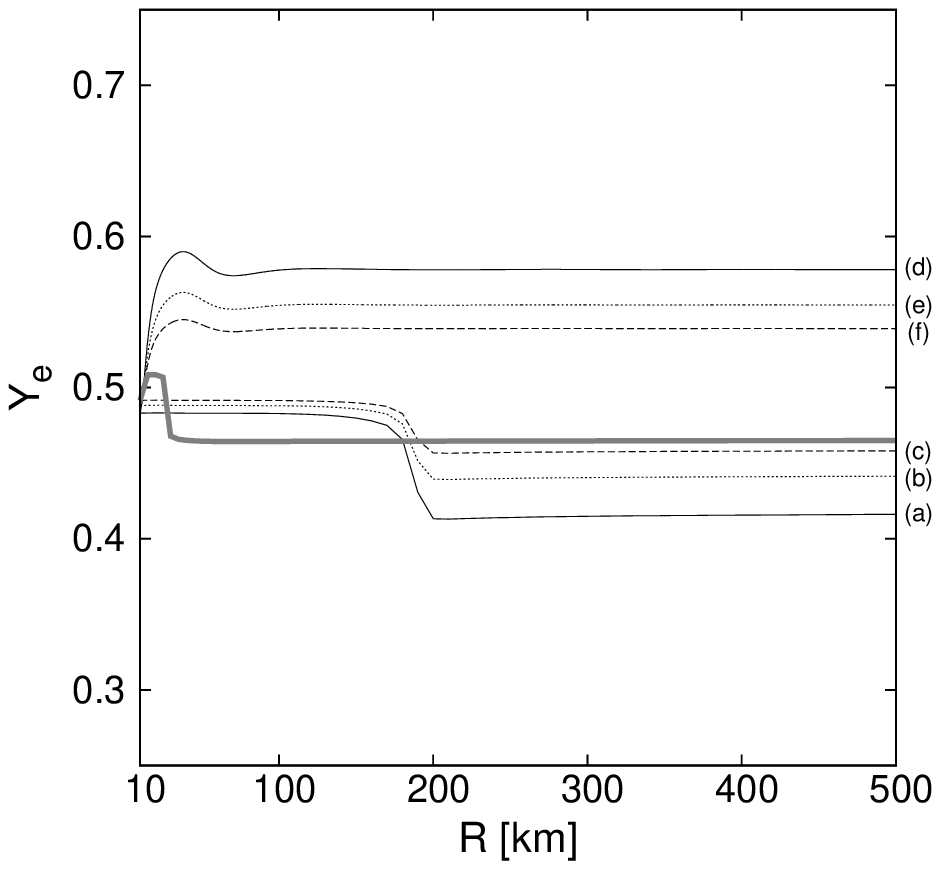, width=350pt}
\caption{$Y_e$ as a function of the radius in the $2+1$-scheme for the late cooling time. The plot was made for $\Delta m_{13}^2=2\times 10^{-3} \, {\rm eV}^2$, $\Delta m_{14}^2=2\,{\rm eV}^2$, $\sin^2 2\theta_{13}=0.09$ and $\sin^2 2\theta_{14}=0.16$. The curves labelled (a), (b) and (c) stand for $X_\alpha=0$, $X_\alpha=0.3$ and $X_\alpha=0.5$ when all the interactions are considered in the Hamiltonian. The curves labelled (d), (e) and (f) stand for $X_\alpha=0$, $X_\alpha=0.3$ and $X_\alpha=0.5$ when only oscillations in vacuum are considered in the Hamiltonian. The calculations were performed using a Fermi-Dirac distribution to describe the initial condition of neutrinos. The thicker line displays the results obtained for $X_\alpha=0.3$ and by using the distribution function of \cite{pllumbi15,tamborra12}.}
\label{figure6}
\end{figure}

In this section we show the effects of the inclusion of sterile neutrinos in a SN-type environment when the sterile neutrino can mix with one of the two active neutrinos, namely the electron-neutrino. The mixing-angle between active-neutrinos considered in this calculation is $\sin^2 2\theta_{13}=0.09$  and the square-mass-difference $\Delta m^2_{13}=2\times 10^{-3} \, {\rm eV}^2$. We assume  a normal-mass-hierarchy for the atmospheric neutrinos \cite{meregaglia16}. In Fig. \ref{figure6} we show the behaviour of the electron fraction as a  function of the radius for the different interactions. This computation were performed mostly for a FD distribution function as initial condition of Eq. (\ref{dif}) (thin lines) and using a power-law distribution function \cite{keil03,tamborra12,pllumbi15} for $X_\alpha=0.3$ (thick line).

\begin{figure}[!h]
\epsfig{figure=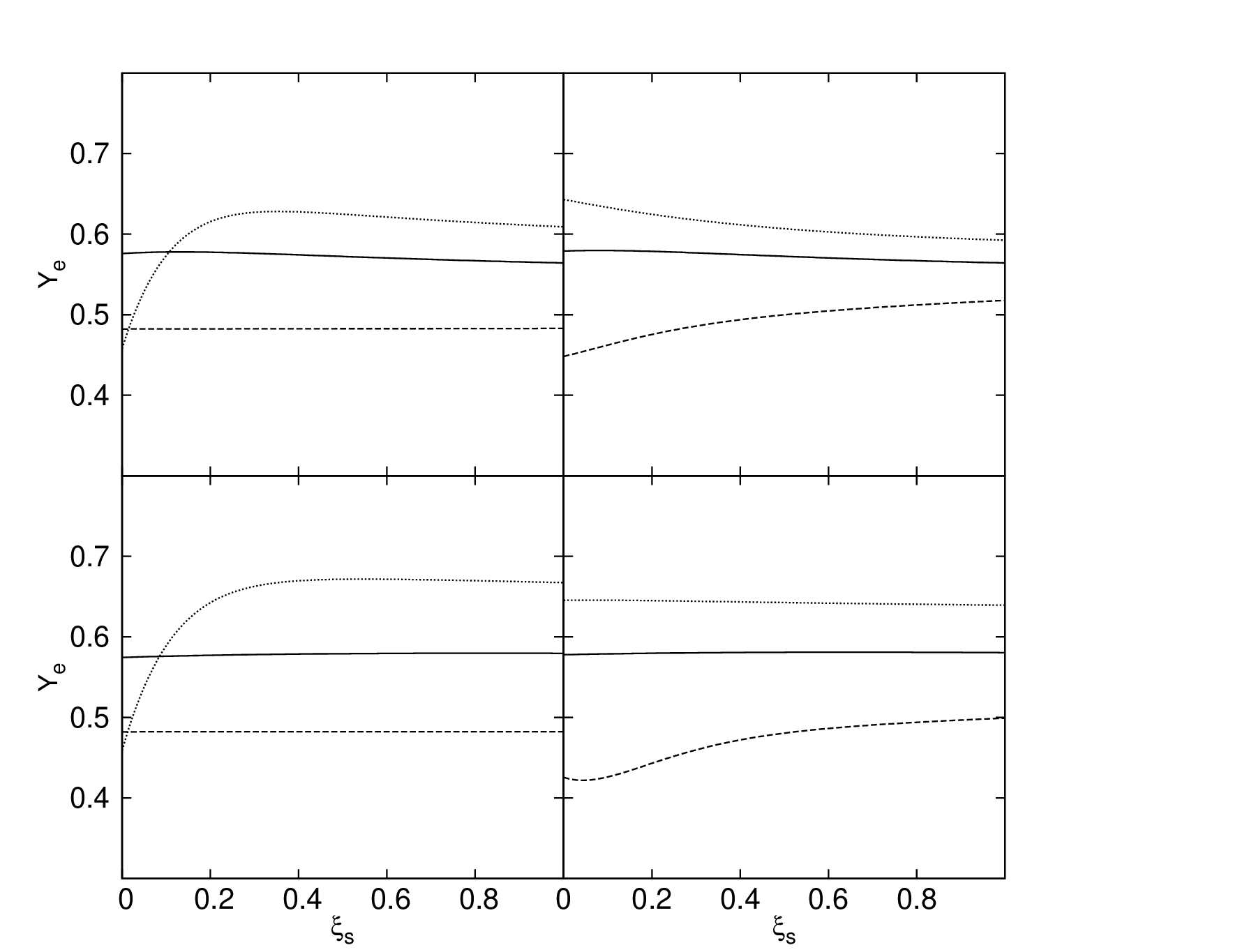,width=250 pt,angle=-90} 
\caption{$Y_e$ as a function of the initial condition $\xi_s$, in the $2+1$-scheme, for two different set of mixing parameters at $\rm{R}= 75\, {\rm Km}$ (first column) and $\rm{R}= 250\, {\rm Km}$ (second column). Solid line: only oscillations; dotted line: oscillations and neutrino-matter interactions and oscillations; dashed-line: neutrino-matter and neutrino-neutrino interactions. Top figure: $\sin^2 2\theta_{14}=0.5$; bottom figure: $\sin^2 2\theta_{14}=0.1$, both sets of results have been obtained with $\Delta m^2_{14}=2 \, {\rm eV}^2$.} 
\label{figure7}
\end{figure}

In Fig. \ref{figure7} we present the results of the electron-fraction  $Y_e$ as a function of the constant $\xi_s$, at $R= 75 \, {\rm Km}$ and $\rm{R}=250 \, {\rm Km}$.
It is observed that the value of the electron-fraction is larger than $0.5$ if the neutrino-neutrino interactions are not included in the calculation. 

\begin{figure}[!h]
\epsfig{figure=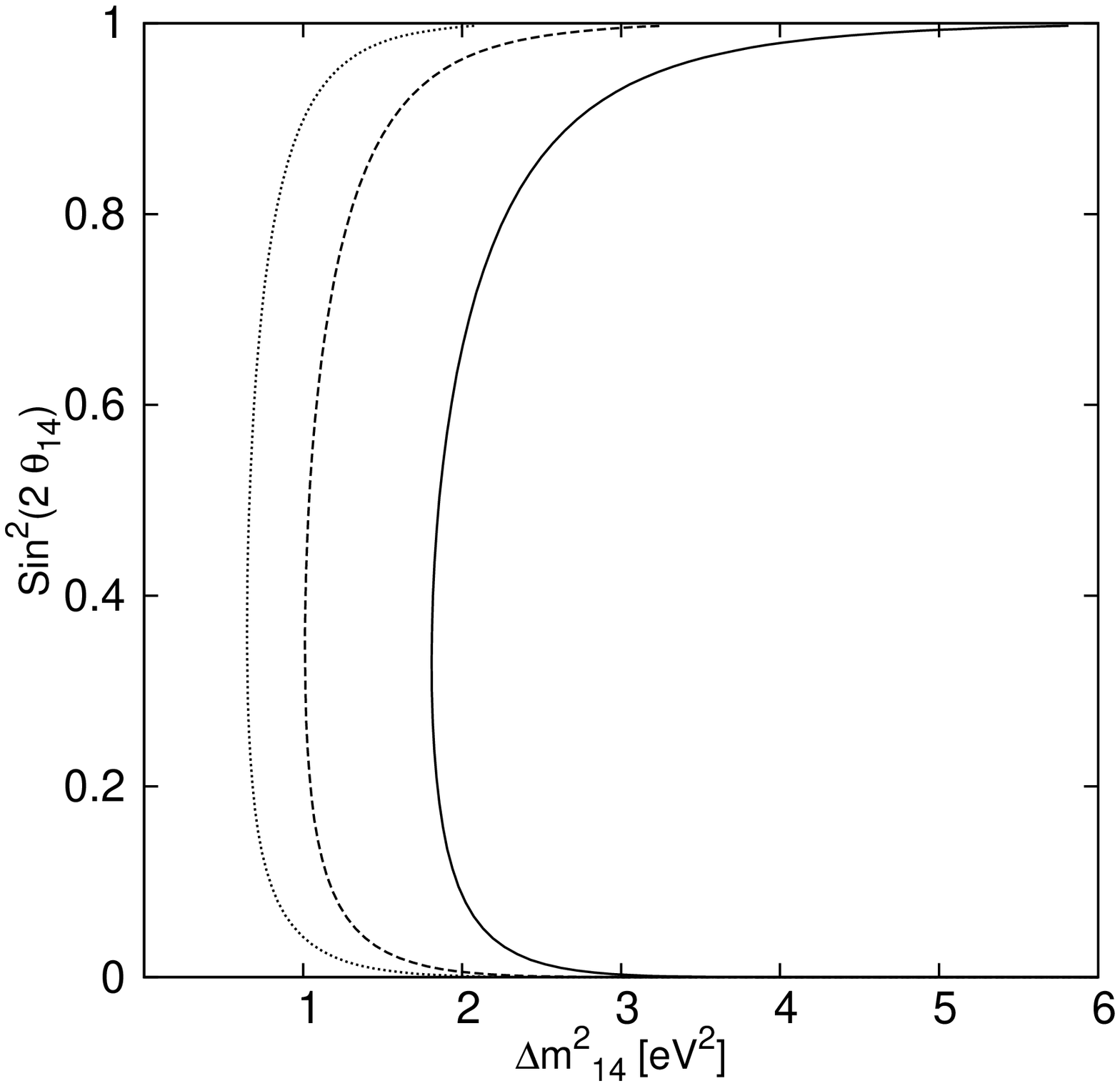,width=250 pt,angle=-90} 
\caption{Parameters consistent with the r-process in the 2+1 scheme. The results are shown in the same way of Figure \ref{figure3}. The lines stand for $Y_e=0.48$, calculated using all the interactions in the neutrino sector. Solid line: $R= 150\, {\rm Km}$; dashed line: $R= 200\, {\rm Km}$; dotted line: $R= 250\, {\rm Km}$.} 
\label{figure8}
\end{figure}

In Fig. \ref{figure8} we show the values of the mixing-angle and the square-mass-difference, between active and sterile-neutrinos, for  $\xi_s=0$ and $Y_e=0.48$. The condition $\Delta m_{14}^2  >2 \, {\rm eV}^2$ is consistent with the r-process at small radius. This value decreases if the radius increases. In the calculations we set  $X_\alpha=0$ since for larger values of $X_\alpha$ we found $Y_e > 0.5$.

\begin{figure}[!h]
\epsfig{figure=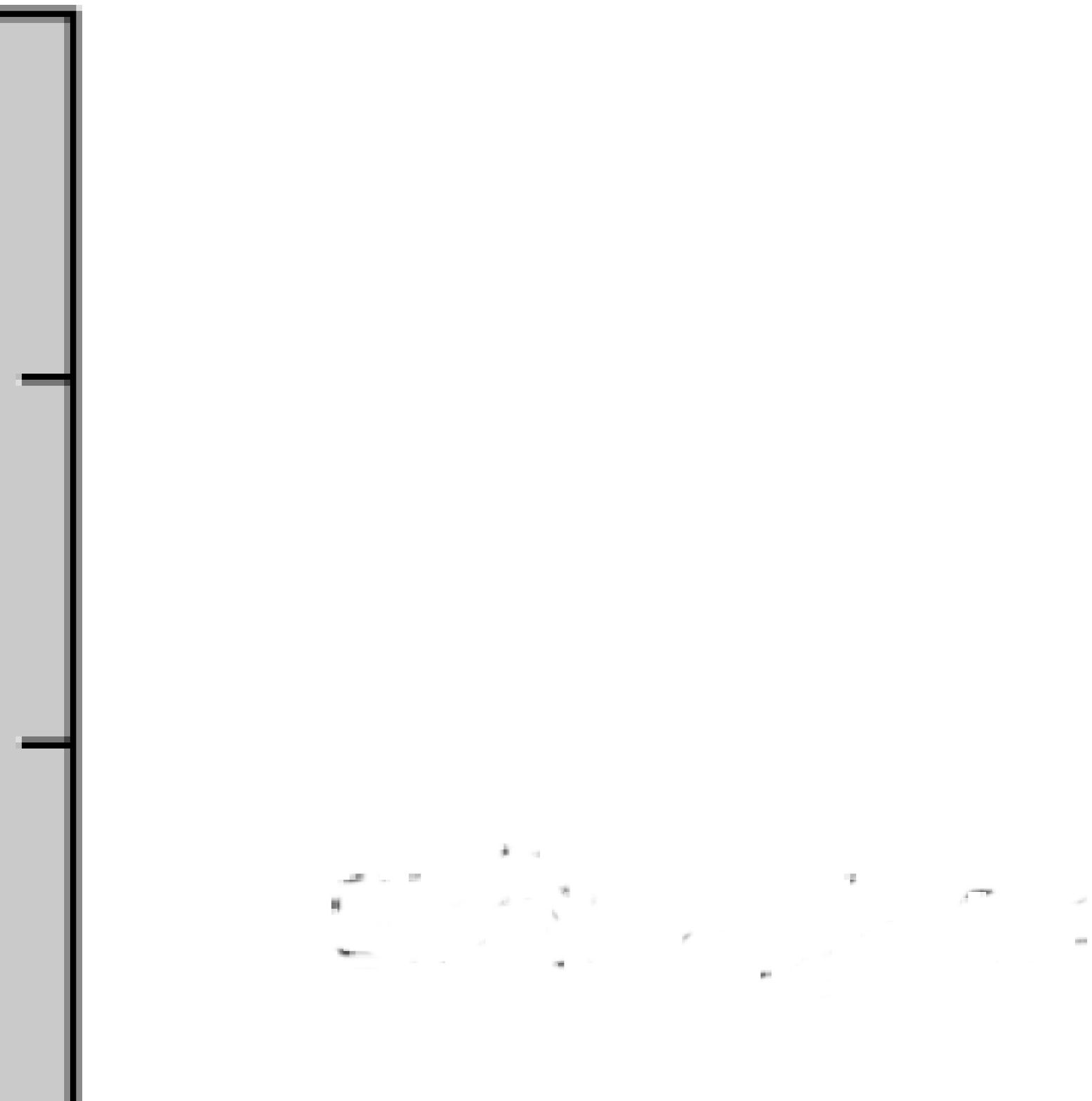,width=400 pt,angle=-180} 
\caption{Allowed values of $\Delta m_{14}^2$ and $\xi_s$ for two different values of the  mixing angle  $\theta_{14}$  and for  $R= 150\, {\rm Km}$ ($2+1$-scheme).  First row corresponds to  $\sin^2 2\theta_{14}=0.1$ and second row for $\sin^2 2\theta_{14}=0.5$. First, second and third column stands for $X_\alpha=0$, $0.3$ y $0.5$ respectively. The curves show the results which obey the constraint $Y_e=0.48$. White regions represent the combinations of parameters favourable for the production r-process, while gray regions correspond to the combinations of parameters that generate $Y_e > 0.48$.} 
\label{figure9}
\end{figure}

In Fig. \ref{figure9} we show the allowed values of $\Delta m_{14}^2$ and $\xi_s$ (white regions) for different values of the active-sterile mixing angle and several values of $X_\alpha$. If $\Delta m_{14}^2 > 1.8 \, {\rm eV}^2$ and $\xi_s <0.45$ the electron fraction becomes lower than $0.48$ for small mixing angles. For larger mixing-angles the constraint $Y_e<0.48$ gives smaller values for the constant $\xi_s$ and does not affect the constraint on the square-mass-difference.

\begin{figure}[!h]
\epsfig{figure=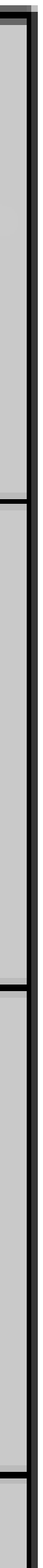,width=400 pt,angle=-180} 
\caption{Allowed values of $\sin^2 2\theta_{14}$ and $\xi_s$ for two values of the  of square-mass-difference $\Delta m_{14}^2$
and for $R= 150\, {\rm Km}$ ($2+1$-scheme). Left figure: $\Delta m_{14}^2 =2 \, {\rm eV}^2$; right figure: $\Delta m_{14}^2 = 10 \, {\rm eV}^2$. White regions represent the combinations of parameters that favors the occurrence of the r-process, while shadowed regions correspond to cases that generate $Y_e > 0.5$. For this case we take $X_\alpha=0$ because for larger values of $X_\alpha$ we have found that $Y_e > 0.5$.}
\label{figure10}
\end{figure}

The constraints on the mixing-angle and the normalization constant of the sterile neutrino distribution functions are more stringent than in the $1+1$-scheme as we can see in Fig. \ref{figure10}. In this case, for $\Delta m_{14}^2 =2 \, {\rm eV}^2$, the mixing angle should be lower than $0.55$ ($\sin^2 2\theta_{14}<0.8$) and $\xi_s <0.1$, in order to obtain $Y_e<0.5$.

As general features of our results we can mention that the inclusion of the sterile neutrino have an important effect upon $Y_e$, since the value of the electron fraction can be drastically reduced when all the channels (oscillations, matter interactions and neutrino interactions) are considered.  The allowed region of the space of parameters favours a small value of the renormalization factor (hindrance factor) $\xi_s$. Also, the allowed regions of parameters are strongly reduced when the value of $X_\alpha$ is increased.

\section{Conclusions}
\label{conclusiones}

In this work we have studied the impact of the inclusion of massive sterile neutrinos upon the physical conditions required for the occurrence of the r-process in  supernovae. The analysis was performed  by calculating the electron-fraction in the stellar interior as a function of the sterile-active neutrino mixing parameters. We have found that the electron abundance is sensitive to the inclusion of sterile neutrinos, and that it depends on the neutrino interactions, e. g. neutrino-neutrino interactions or neutrino-matter interactions. From our results, the onset of the r-process in presence of a sterile neutrino, is compatible with the limits $\Delta m_{14}^2 \geq 2 {\rm eV}^2$, $\sin^2 2\theta_{14}<0.8$, and $\xi_s <0.5$, for the square-mass-difference, mixing-angle and enhancement  factor of the sterile neutrino sector, respectively. As explained in the text, we found that the r-process is strongly affected by the fraction $X_{\alpha}$.


\section*{{\bf Acknowledgment}}
This work was supported by  grant (PIP-616) of the National Research Council of Argentina (CONICET), and by a research-grant of the National Agency for the Promotion of Science and Technology (ANPCYT) of Argentina. O. C. and M. E. M. are members of the Scientific Research Career of the CONICET.


\end{document}